\documentclass[prl,aps,twocolumn]{revtex4}
\usepackage[latin1]{inputenc}
\usepackage{amsmath}
\usepackage{amsfonts}
\usepackage{amssymb}
\usepackage{graphicx}
\newcommand{\etal}{{\em{et al.}}}
\newcommand{\eg}{{\it e.g.}}
\newcommand{\ie}{{\it i.e.}}
\newcommand{\beq}{\begin{equation}}
\newcommand{\eeq}{\end{equation}}
\def\lsim{\mathrel{\rlap{\lower4pt\hbox{\hskip1pt$\sim$}}
    \raise1pt\hbox{$<$}}}         
\def\gsim{\mathrel{\rlap{\lower4pt\hbox{\hskip1pt$\sim$}}
    \raise1pt\hbox{$>$}}}         
\begin{document}
\title{In-Medium Effects on Charmonium Production in Heavy-Ion Collisions}
\author{
Lo\"{\i}c Grandchamp$^{1,2}$, 
Ralf Rapp$^{3,4}$ and 
Gerald E. Brown$^{1}$ } 
\affiliation{
$^{1}$Department of Physics, SUNY Stony Brook, NY 11794-3800\\
$^{2}$IPN Lyon, 69622 Villeurbanne Cedex, France\\
$^{3}$NORDITA,$\;\!$Blegdamsvej$\,\,\!$17,$\;\!$2100$\;\!$Copenhagen,$\;\!$Denmark\\
$^{4}$Cyclotron Institute and Physics Department, Texas A\&M University,
      College Station, TX 77843
}
\date{\today}
\begin{abstract}
Charmonium production in heavy-ion collisions is investigated within 
a kinetic theory framework incorporating in-medium properties of open- 
and hidden-charm states in line with recent QCD lattice calculations.  
A continuously decreasing open-charm threshold across the phase 
boundary of hadronic and quark-gluon matter is found to have important 
implications for the equilibrium abundance of charmonium states. 
The survival of $J/\psi$ resonance states above the transition 
temperature enables their recreation also in the Quark-Gluon Plasma.
Including effects of chemical and thermal off-equilibrium, 
we compare our model results to available experimental data at 
CERN-SPS and BNL-RHIC energies. In particular, earlier found 
discrepancies in the $\psi'/\psi$ ratio can be resolved.
\end{abstract}
\maketitle

The production systematics of heavy-quark bound states in (ultra-) 
relativistic collisions of heavy nuclei ($A$-$A$) is believed to 
encode valuable 
information on the hot and dense strong-interaction matter formed in 
these reactions~\cite{Vogt99}. Based on the notion that charm quark 
pairs ($c\bar c$) are exclusively created in primordial (hard) 
nucleon-nucleon ($N$-$N$) collisions, it has been suggested~\cite{MS86} 
that a suppression of observed $J/\psi$ mesons in sufficiently central 
and/or energetic $A$-$A$ reactions signals the formation of a 
deconfined medium (Quark-Gluon Plasma=QGP), as tightly bound 
$c\bar c$ states are conceivably robust in ha\-dronic matter. 
While theoretical~\cite{LMW95} 
and (indirect) experimental evidence~\cite{phenix-cc} supports 
$N$-$N$ collision-scaling of total charm production, it has 
recently been realized~\cite{pbm00,TSR01,Gor01} that coalescence 
of $c$ and $\bar c$ quarks can induce significant
regeneration of charmonium states in later stages of $A$-$A$ 
collisions, especially if several pairs are present
(\eg, $N_{c\bar c}$=10-20 in central $Au$-$Au$ at RHIC). 
This is a direct consequence of (elastic) $c$-quark reinteractions,  
facilitating the backward direction of charmonium dissociation
reactions, 
$J/\psi + X_1\rightleftharpoons X_2 + c + \bar{c} \ (D+\bar{D})$.
  
Recent lattice computations of Quantum Chromodynamics (QCD) have 
revealed important information 
on charm(onium) properties at finite temperature $T$, most notably: 
{\em{(i)}} an in-medium reduction of the open-charm threshold which 
is surprisingly continuous even across the phase transition 
region~\cite{KLP01}, and,
{\em{(ii)}} the survival of $J/\psi$ and $\eta_c$ states as 
resonances in the QGP phase~\cite{DKP02,UNM02}, with essentially 
unmodified masses. 
Note that the use of charmonium spectral functions, 
in connection with appropriate asymptotic states, incorporates both 
(static) screening and (dynamical) dissociation mechanisms.

In this letter, we propose an approach that implements charm properties 
inferred from lattice QCD as microscopic in-medium effects into a 
kinetic rate equation.  It enables a comprehensive treatment of 
charmonium dissociation and regeneration across the phase transition, 
connecting hadronic and QGP phases in a continuous way, not present
in previous calculations. {\it E.g.}, in the kinetic approach of 
Ref.~\cite{TSR01}, $J/\psi$'s were only considered in the QGP phase 
using their vacuum binding energy, 
whereas recent transport calculations~\cite{ZKL02,BCS03} 
do not invoke the notion of a (equilibrium) phase transition.  
In addition, in-medium open-charm masses will significantly affect
the equilibrium levels of charmonium, as employed in statistical 
models~\cite{pbm00,Gor01}. 
We thus conceptually improve our earlier construc\-ted ``two-component'' 
model~\cite{GR01}, which combined statistical production at 
hadronization with suppression in hadronic and QGP phases. 
Employing a schematic thermal fireball model, the 
rate equations are applied to heavy-ion reactions at SPS and RHIC.  

We begin by evaluating in-medium masses of charm states. 
Charmonium masses appear to be essentially un\-affected at finite 
$T$, even above $T_c$ (critical temperature)~\cite{DKP02,UNM02}, 
which we assume from now on. Un\-quenched QCD lattice studies of the free 
energy, $F_{Q\bar{Q}}(r,T)$, of a heavy-quark pair~\cite{KLP01}, 
however, exhibit a plateau value for $r\to\infty$ which gradually 
decreases with increasing $T$. In the hadron gas (HG), this can be 
interpreted as a reduction of $D$-meson masses (supported by QCD 
sum rules~\cite{Hay00}) being driven by a reduced constituent 
light-quark mass, $m_q^*$, due to (partial) chiral symmetry 
restoration~\cite{DPS01}.  
We implement this also for other open-charm 
hadrons by employing a Nambu-Jona-Lasinio model calculation for 
$m_q^*$ at finite $T$ and quark chemical potential $\mu_q$~\cite{BMZ87},
neglecting heavy-light quark interaction and kinetic energies.  
For small $\mu_q$, the typical light-quark mass reduction amounts to 
$\Delta m_q (T_c)$$\simeq$140~MeV.  To assess open-charm states in 
the QGP, a key element is the essentially continuous behavior
of the open-charm threshold through $T_c$. Within a quark description 
above $T_c$, we propose to identify this with an in-medium 
charm-quark mass, $m_c^*$$\simeq$1.6-1.7~GeV~\cite{RG03}.   
In fact, $m_c^*(T_c)$ is fixed such that the {\em total} open-charm 
density smoothly matches the hadronic side.
The difference to the bare charm-quark mass  ($m_c$$\simeq$1.3~GeV)
is naturally attributed to a thermal correlation energy of heavy quarks 
in the QGP.  

Next, we determine the thermal widths (or inverse lifetimes), 
$\Gamma_\psi$=$(\tau_\psi)^{-1}$, of charmonium states $\psi$. They 
are related to (inelastic) dissociation cross sections, 
$\sigma^{diss}_{\psi i}$, by  
\beq
\Gamma_\psi(T) = \sum_{i} \int \frac{d^3k}{(2\pi)^3} \ f^i(\omega_k;T) 
 \  \sigma^{diss}_{\psi i} \ v_{rel}  \ ,   
\label{eq:tau}
\eeq
where $i$ runs over all matter constituents with 
thermal distributions $f^i$.  Within the QGP, parton-induced 
``quasifree'' break-up, $i + \psi \to i+c+\bar c$ 
($i$=$g$,$q$,$\bar q$), is utilized~\cite{GR01}. For small 
binding energies, $E_B$=$2m_c^*$-$m_\psi$$\lsim$$\Lambda_{QCD}$,
the latter (with $\alpha_s$$\simeq$0.25) has been shown~\cite{GR01}
 to dominate standard gluo-dissociation~\cite{Shu78},
$g + \psi \to  c+ \bar c$.  In addition,  
essentially unbound charmonium resonances such as $\psi'$ 
and $\chi$-states can be treated on an equal footing.   
The thermal $J/\psi$ width turns out to be  
$\sim$100~MeV at $T$$\simeq$250~MeV, reminiscent to 
(quenched) lattice results~\cite{UNM02} under comparable conditions.   
  
For $J/\psi$'s in the HG, we compute inelastic cross sections
with $\pi$-~\cite{Dura02} and $\rho$-mesons
within a flavor-$SU(4)$ ef\-fective lagrangian formalism~\cite{LK00,HG01}, 
including in-medi\-um charm hadron mas\-ses~\cite{Sib99} as described 
above. The free $\chi$ and $\psi'$ cross sections are approximated 
by geometric scaling~\cite{GR02}, 
supported by quark-model calculations~\cite{WSB01}. If the 
$D\bar D$ threshold moves below the charmonium mass, we additionally 
evaluate direct decays, $\psi\to D\bar D$, accounting for 
wavefunction-node effects according to Ref.~\cite{FLS02}.
Despite the increase in available phase space due to reduced
masses, the charmonium 
lifetimes in the HG remain longer than in the QGP.  

\begin{figure}
\includegraphics[width=0.44\textwidth,clip=]{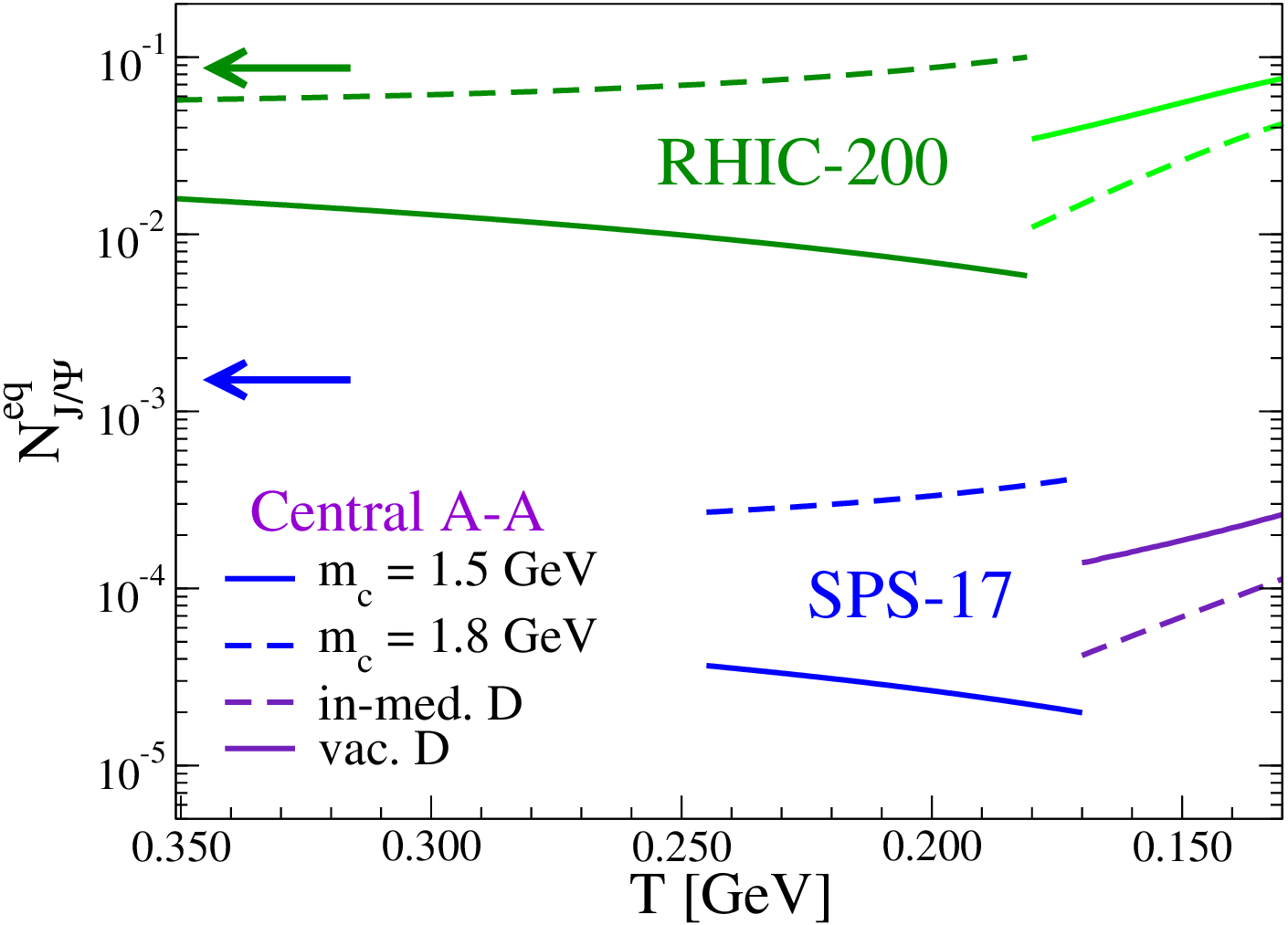}
\vspace*{-0.15cm}
\caption{Equilibrium $J/\psi$ abundances (with 
$m_\psi(T)\equiv m_\psi^{vac}$) in an isotropic, adiabatically expanding 
system at fixed $N_{c\bar c}$ for SPS and RHIC conditions for two 
values of  the charm-quark mass above $T_c$, and for free and in-medium 
charmed hadron masses below $T_c$.
}
\label{fig:jpsi_eq}
\end{figure}

In-medium charm properties have important 
consequences for charmonia in heavy-ion reactions.  
To illustrate this point, consider the $J/\psi$ number in 
thermal equilibrium, $N_{\psi}^{eq}$=$V_{FB} n_{\psi}^{eq}$
($V_{FB}$: 3-volume), with the density
\beq
n_{\psi}^{eq}(T,\gamma_c) = 3 \gamma_c^2\int
 \frac{d^3q}{(2\pi)^3}f^{\psi}(m_{\psi},T) \ . 
\label{eq:npsi}
\eeq 
Here, we have included chemical off-equilibrium effects via the
charm-quark fugacity $\gamma_c$ which is adjusted to the total
number, $N_{c\bar c}$, of $c\bar c$ pairs in the system 
via~\cite{pbm00,Gor01} 
\beq
N_{c\bar c}=\frac{1}{2} \gamma_c N_{op}
\frac{I_1(\gamma_c N_{op})}{I_0(\gamma_c N_{op})}+
V_{FB}\sum\limits_{\eta_c,J/\psi,\dots}n_\psi(T,\gamma_c)
\label{eq:Ncc}
\eeq 
where $N_{op}$=$V_{FB} n_{op}(m_{c,D}^*;T)$ denotes the total equilibrium-number 
of open-charm states in the thermal fireball.  
This procedure resides on the expectation 
that essentially all charm-quarks are created in primordial 
$N$-$N$ collisions, \ie,  
$N_{c\bar c}$ does not chemically equilibrate~\cite{LMW95}.  
Thermal equilibration of charm quarks, however, is 
conceivable~\cite{Svet88}, implying a hierarchy of relaxation
times as 
$\tau_{c,\bar c}^{chem}\gg\tau_{FB} \gg \tau_{c,\bar c}^{therm}$.  
The resulting $J/\psi$
equi\-librium abundances, $N_{\psi}^{eq}$, are shown in Fig.~\ref{fig:jpsi_eq}
under conditions resembling central 
$Pb$-$Pb$ ($Au$-$Au$) collisions at SPS (RHIC).
One observes a large sensitivity to the (in-medium) open-charm masses.
At fixed $N_{c\bar c}$, larger values for $m_c^*$ (or $m_D^*$)  imply 
a thermal suppression of open-charm states so that an increasing 
number of anti-/charm quarks is 
redistributed into charmonia ($c\bar c$ states).   
Enforcing continuity on the open-charm spectrum across $T_c$ then
has the interesting consequence that, due to the volume increase
in the hadronic phase (reducing $\gamma_c$), the equilibrium charmonium 
level on the QGP side is significantly {\em larger} than
on the HG side~\cite{RG03}, \ie,
$J/\psi$ formation is {\em favored} in the QGP.

Also note that, due to a constant $N_{c\bar c}$, the equilibrium 
$J/\psi$ numbers in the hadronic phase {\em grow} with decreasing 
temperature. Thus, if $J/\psi$'s equilibrate close to $T_c$, 
subsequent hadronic reactions will tend to increase their 
abundance (which has indeed been found in transport calculations
at RHIC energies~\cite{ZKL02,BCS03}), quite contrary  
to the commonly assumed hadronic dissociation.

To model the time dependence, $N_\psi(\tau)$, of charmonia in 
heavy-ion collisions, we utilize the dissociation (formation) time 
$\tau_\psi$, Eq.~(\ref{eq:tau}), and equilibrium density, 
Eq.~(\ref{eq:npsi}), within a kinetic rate equation,   
\beq
\frac{dN_{\psi}}{d\tau} =
-\frac{1}{\tau_{\psi}}\left[N_{\psi}-N_{\psi}^{eq}\right] \ . 
\label{eq:rate}
\eeq 
The underlying temperature and volume evolution, $T(\tau)$ and 
$V_{FB}(\tau)$, are taken from a (schematic) thermal fireball expansion 
which is consistent with observed hadro-chemistry and radial flow 
characteristics (as employed before for thermal 
dilepton production~\cite{RW00}). 

The simple form of the gain term in 
Eq.~(\ref{eq:rate}) resides on the assumption that 
the surrounding light and open-charm constituents (quarks or
hadrons) are in thermal equilibrium. Following 
Ref.~\cite{GR02}, we relax this assumption for $c$-quarks by 
introducing a thermal relaxation time correction, 
$\mathcal{R}=1-\exp(-\int d\tau/\tau_{c}^{therm})$, which reduces
$N_{\psi}^{eq}$ in the early phases. Varying $\tau_{c}^{therm}$
within a factor of 2, the thermal yield is affected by 
$\pm 10\%$.

Another correction concerns the effective volume over which the charm
quantum number is conserved (figuring into Eq.~(\ref{eq:Ncc}) via 
the Bessel functions). Clearly, if only few
$c\bar c$ pairs are present, their pointlike primordial production 
implies that they cannot explore the entire fireball volume in the 
early stages\footnote{We thank U. Heinz for pointing this out to us, 
including useful discussions on the correlation volume.}. 
This problem is well-known from strange particle production at fixed 
target energies, where a phenomenological ``correlation volume'', $V_0$, 
has been introduced to {\em localize} strangeness 
conservation~\cite{HRT00}. We adopt the same procedure here 
for local charm conservation by replacing 
$V_{FB}(\tau)$ in the argument of the Bessel functions in 
Eq.~(\ref{eq:Ncc}) with 
$V_0(\tau)$=$4\pi(r_0 + \langle v_c\rangle \tau)^3/3$.  
$r_0$$\simeq$1.2~fm represents a minimal radius characterizing the
range of strong interactions, and $\langle v_c\rangle$$\simeq$0.5$c$
is the average relative speed of produced $c$ and $\bar c$ 
quarks as inferred from experimental $D$-meson 
$p_\perp$-distributions~\cite{WA92-97}. 
We checked that our results are not very sensitive 
to the parameterization for $V_0$ (\textit{e.g.},
varying $\langle v_c\rangle$ within a factor of 2 requires
a change in the effective coupling constant $\alpha_s$ by 
$\pm 30\%$, to fit SPS data).


The rate Eq.~(\ref{eq:rate}) is readily integrated over the fireball 
evolution of the collision for $J/\psi$, $\psi'$ and $\chi$'s,  
once initial conditions are specified. 
In an $A$-$A$ reaction at given centrality, $N_{c\bar c}$
is determined by the collision-scaled $N$-$N$ cross section.
The initial charmonium numbers are also assumed to follow hard 
production (\ie, empirical fractions of $N_{c\bar c}$ observed in
$N$-$N$ collisions), subject to (pre-equilibrium) nuclear absorption 
with a recently updated absorption cross section 
$\sigma_{abs}$=4.4~mb~\cite{NA50-03}.   
Typical values for the fireball thermalization time range from 
$\frac{1}{3}$ (RHIC-200) to 1~fm/c (SPS).

We first confront our approach to   
NA50/NA38 data in $\sqrt{s_{NN}}$=17.3~GeV $Pb$-$Pb$ at 
CERN-SPS. 
\begin{figure}
\includegraphics[width=0.42\textwidth,clip=]{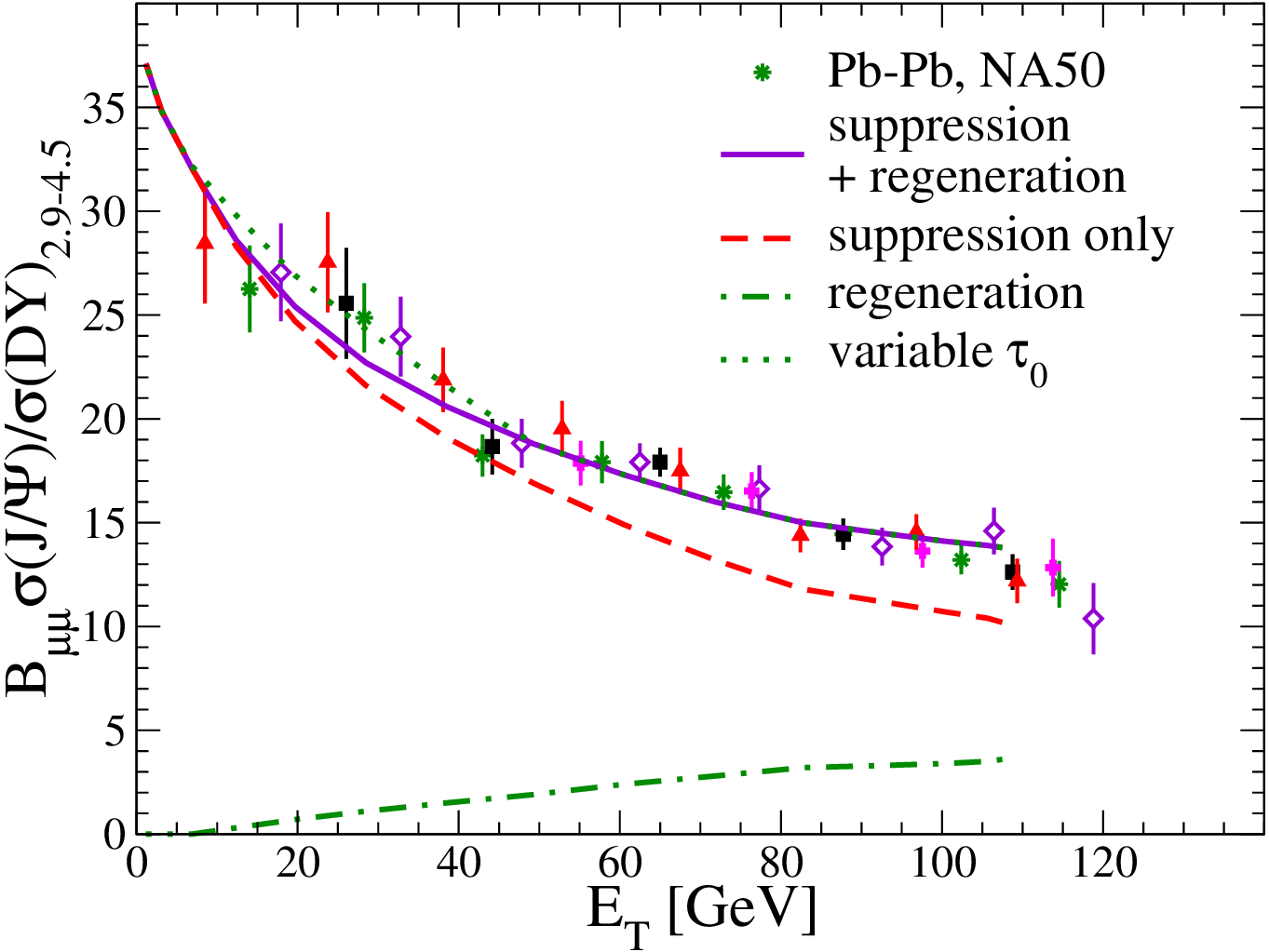}
\vspace*{-0.15cm}
\caption{Centrality dependence of $J/\psi$/Drell-Yan dimuons  
 at SPS; NA50 data~\cite{NA50-03} are compared to our results
 with (solid line) and without (dashed line) charmonium regeneration 
 (represented by the dot-dashed line). The dotted line
 includes longer thermalization times $\tau_0$ in peripheral 
 collisions.}
\label{fig:jpsi-sps}
\end{figure}
Fig.~\ref{fig:jpsi-sps} displays the ratio of 
$J/\psi$$\to$$\mu\mu$ to Drell-Yan di\-muons as a function of centrality.  
The agreement be\-tween model
(solid line) and data is fair for semi-/central 
collisions (at $E_T$$>$100~GeV, the data can be reproduced by 
accounting for transverse energy fluctuations and losses in the minimum 
bias analysis at impact parameters close to zero~\cite{CKS02}, 
cf.~Ref.~\cite{GR02}).
Since the initial $J/\psi$ number is well above the equilibrium level 
(cf.~lower arrow in Fig.~\ref{fig:jpsi_eq}), 
$J/\psi$ regeneration (the gain term in Eq. (\ref{eq:rate})) is
very moderate (dot-dashed line). Therefore, in line with our previous 
findings~\cite{GR01}, $J/\psi$ {\em suppression} is the main effect 
at SPS energies.   

In peripheral collisions, the suppression appears to be
slightly overestimated. We believe this discrepancy to reside in
limitations of our fireball description. In particular, thermalization
is expected to be delayed (and/or incomplete) at large impact 
parameters due to less energetic initial conditions. This is also borne 
out of hydrodynamic models, which, \eg, reproduce the observed 
elliptic flow for mid-central collisions, but overestimate it for 
peripheral ones. A suitable increase of the equilibration time 
by up to a factor of $\sim$3~\cite{Ko03} indeed
improves the agreement at small $E_T$, cf.~dotted line in 
Fig.~\ref{fig:jpsi-sps}.

\begin{figure}
\includegraphics[width=0.44\textwidth,clip=]{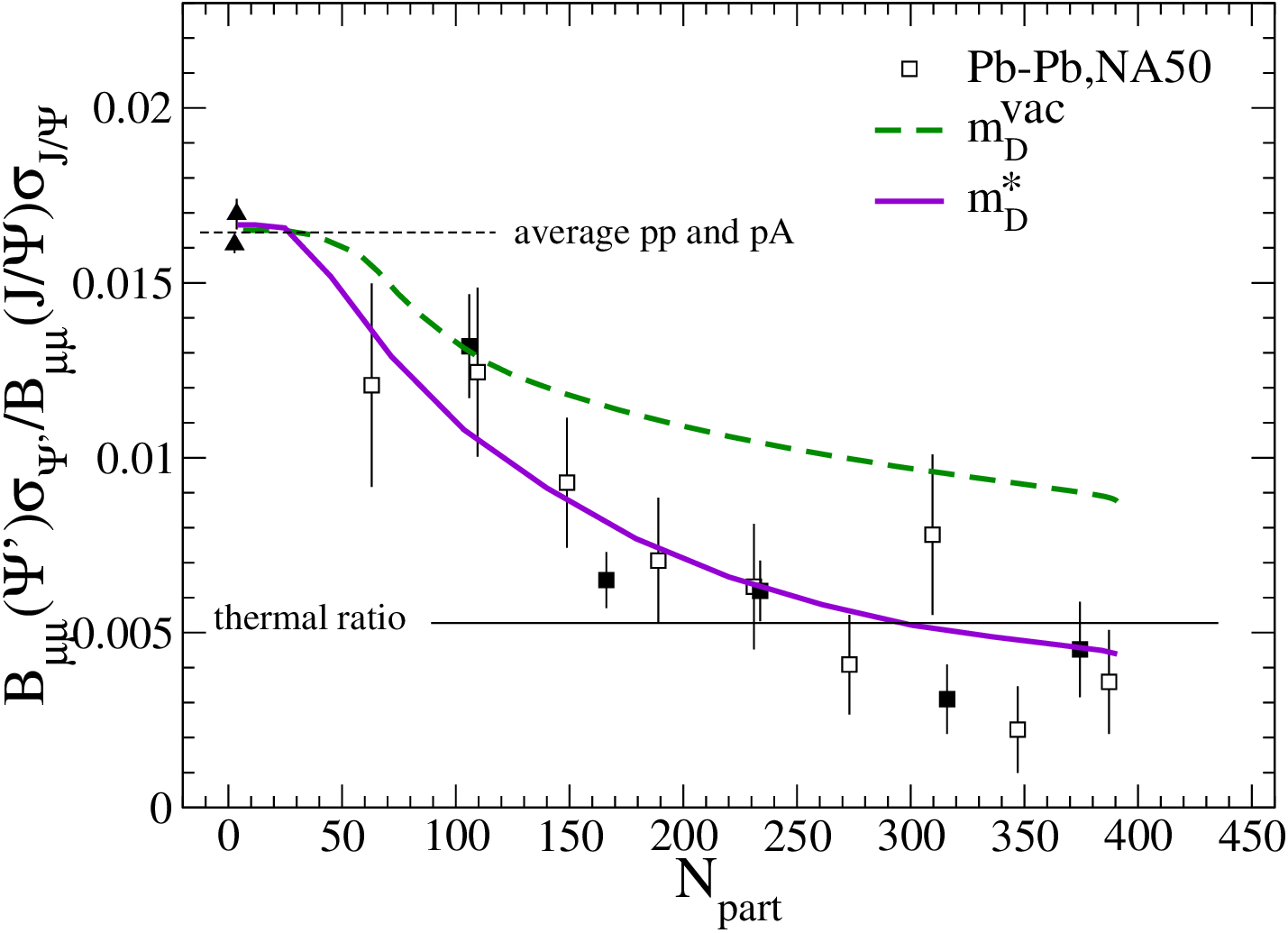}
\vspace*{-0.15cm}
\caption{Centrality dependence of the $\psi'/\psi$ data~\cite{NA50-psip} 
at SPS compared to our results  with (full line) and without (dashed line) 
in-medium reduced $D$-meson masses. 
}
\label{fig:psip-psi-sps}
\end{figure}

Another important observable is the $\psi'/\psi$ ratio. In
Ref.~\cite{GR02}, the $\psi'$ dissociation  
rates were too small by a factor of $\sim$5 to account for NA50 
data~\cite{NA50-psip}. With in-medium $D$-meson masses, however, 
$\Gamma_{\psi'}^{had}$ increases substantially, primarily due to 
the opening of the $\psi'\to D\bar D$ decay channel. As a result, 
the $\psi'/\psi$ data are reasonably well described, 
cf.~Fig.~\ref{fig:psip-psi-sps}.  

\begin{figure}
\includegraphics[width=0.44\textwidth,clip=]{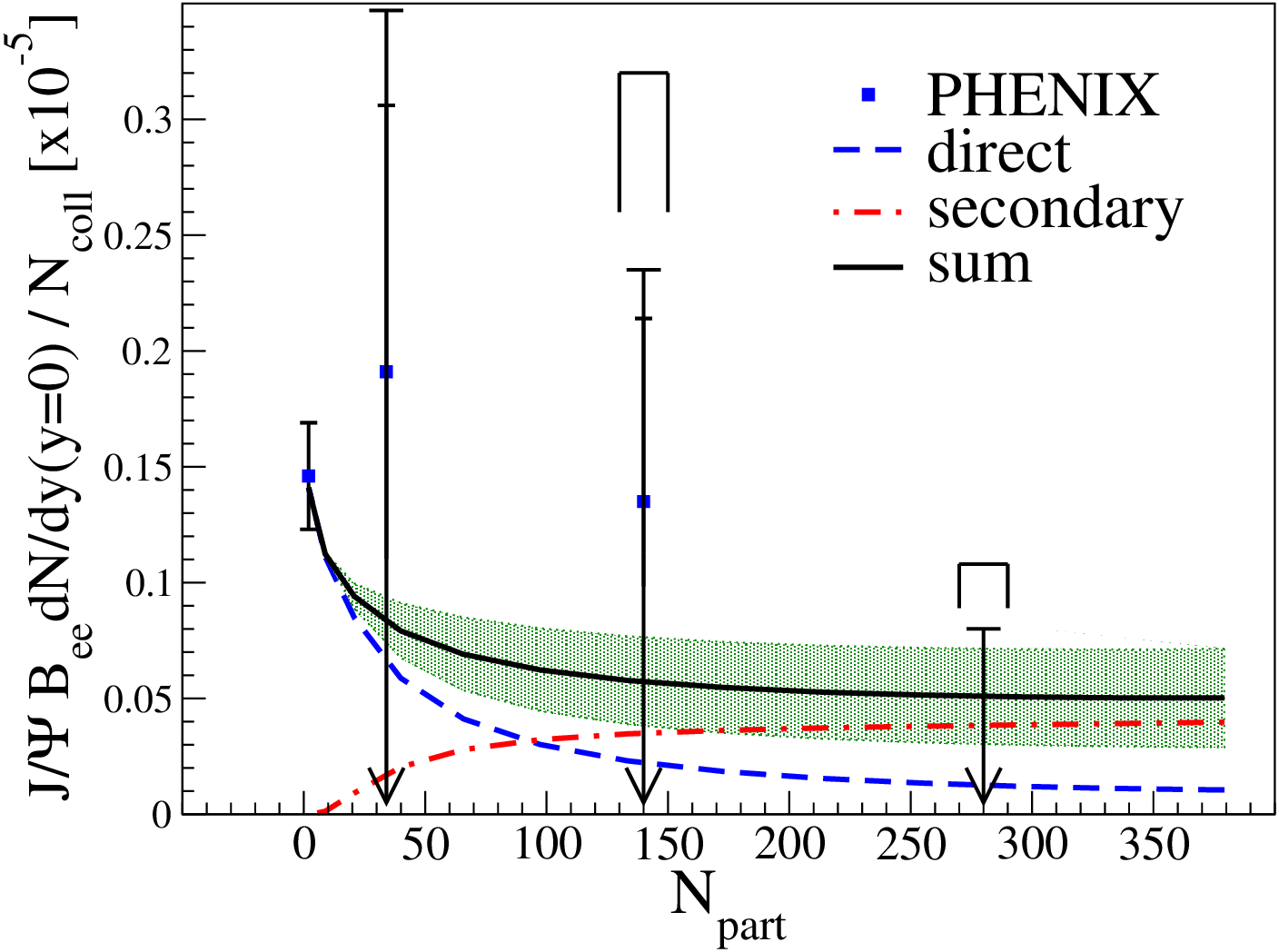}
\vspace*{-0.15cm}
\caption{$J/\psi$ yield per binary $N$-$N$ collision versus 
participant number in $\sqrt{s_{NN}}$=200~GeV $Au$-$Au$ collisions. 
Preliminary PHENIX data~\cite{PHE-03} are compared to our model
calculations; dot-dashed line: thermal regeneration; dashed line: 
suppressed primordial production; band: total $J/\psi$ yields with 
different values for in-medium open-charm masses.}
\label{fig:jpsi-rhic}
\end{figure}

We finally examine the impact of in-medium modifications 
at RHIC, cf.~Fig.~\ref{fig:jpsi-rhic}. 
Since reduced $D$-meson masses entail a lower $J/\psi$ equilibrium level
(cf.~Fig.~\ref{fig:jpsi_eq}), the regeneration of $J/\psi$'s 
is somewhat less pronounced than the statistical production with free 
hadron masses in the 2-component model~\cite{GR02,GR03}. 
Nevertheless, in central $Au$-$Au$ collisions, regenerated $J/\psi$'s  
still exceed the suppressed primordial contribution, with the 
total yield (solid curve) in line  
with first PHENIX data~\cite{PHE-03}. The uncertainty in the 
charm-hadron mass reduction is illustrated by the band in
Fig.~\ref{fig:jpsi-rhic}, corresponding to 
80~MeV$<$$\Delta m_q(T_c)$$<$250~MeV 
with accordingly matched charm-quark masses in the QGP.     
In any case, most of the regeneration occurs above $T_c$.

To summarize, we have proposed a conceptually improved model of 
charmonium production 
in heavy-ion collisions. In-medium modifications of both open-charm
and charmonium states have been modeled in accordance with  
recent finite-$T$ QCD lattice calculations. The apparent reduction of 
the open-charm threshold with increasing $T$ has been linked to (partial) 
chiral symmetry restoration in the $SU(2)$-sector via 
decreasing constituent light-quark masses in charmed hadrons. The 
continuity of the open-charm threshold across the phase transition has 
been encoded in thermal charm-quark masses in the QGP phase, 
together with $T$-independent charmonium masses and $J/\psi$ resonance
states surviving above $T_c$. 
Upon inclusion of chemical and thermal off-equilibrium effects, we solved
kinetic rate equations for the time evolution of charmonia in
heavy-ion reactions. The results for $J/\psi$ centrality dependences 
compare well with data from SPS and RHIC. The earlier predicted 
transition from suppressed production at SPS to 
predominant (thermal) regeneration at RHIC persists, with both 
mechanisms residing on QGP formation. A much
improved description of the $\psi'/\psi$ ratio at SPS emerged due to 
hadronic in-medium effects, accelerating $\psi'$ dissociation. 
Forthcoming measurements by NA60 (SPS) and PHENIX (RHIC) will
be of great importance to scrutinize the proposed approach.  

{\it Acknowledgments}: We thank U.~Heinz and E.V.~Shu\-ryak
for valuable discussions. L.G. is grateful for the hos\-pitality 
at NORDITA and the Niels-Bohr Institute where part of this work was
done within a Rosenfeld fellowship. This work was supported in 
part by the US Department of Energy under Grant No. DE-FG02-88ER40388.

\def\IJMPA{{Int. J. Mod. Phys.} {\bf A}}
\def\EPJ{{Eur. Phys. J.} {\bf C}}
\def\JPG{{J. Phys} {\bf G}}
\def\JHEP{{J. High Energy Phys.}}
\def\NPA{{Nucl. Phys.} {\bf A}}
\def\NPB{{Nucl. Phys.} {\bf B}}
\def\PLB{{Phys. Lett.} {\bf B}}
\def\PLC{Phys. Repts. }
\def\PRL{Phys. Rev. Lett. }
\def\PRD{{Phys. Rev.} {\bf D}}
\def\PRC{{Phys. Rev.} {\bf C}}
\def\ZPC{{Z. Phys.} {\bf C}}

\end{document}